\documentclass{PoS}

\title{On the $B^{*'}\to B$ transition}

\ShortTitle{On the $B^{*'}\to B$ transition}

\author{Beno\^it~Blossier\\
        Laboratoire de Physique Th\'eorique (B\^at. 210), Universit\'e Paris Sud,  Centre d'Orsay, 91405 Orsay-Cedex, France }

\author{\speaker{Antoine~G\'erardin}\\
        Laboratoire de Physique Th\'eorique (B\^at. 210), Universit\'e Paris Sud,  Centre d'Orsay, 91405 Orsay-Cedex, France\\
        E-mail: \email{antoine.gerardin@th.u-psud.fr}}
        
\author{John~Bulava\\
        School of Mathematics, Trinity College, Dublin 2, Ireland }
        
\author{Michael~Donnellan\\
DESY, Platanenallee 6, D-15738 Zeuthen, Germany}
        
\abstract{
We present a first $N_{\rm f}=2$ lattice estimate of the hadronic coupling 
$g_{12}$ which parametrizes the strong decay of a radially excited $B^*$ meson 
into the ground state $B$ meson at zero recoil. We work in the static limit of  Heavy Quark Effective Theory (HQET) and solve  
a Generalised Eigenvalue Problem (GEVP), which is necessary for the extraction 
of excited state properties. After an extrapolation to the continuum limit 
and a check of the pion mass dependence, we obtain $g_{12} = -0.17(4)$. 
}

\FullConference{31st International Symposium on Lattice Field Theory - LATTICE 2013\\
		July 29 - August 3, 2013\\
		Mainz, Germany}

\newcommand{\fm}{{\rm fm}}
\newcommand{\mev}{{\rm MeV}}
\newcommand{\Vcal}{{\mathcal{V}}}
\newcommand{\Acal}{{\mathcal{A}}}
\newcommand{\Pcal}{{\mathcal{P}}}

\usepackage{amsmath}
\usepackage{amssymb}

\usepackage{fontenc}
\usepackage{times}
\usepackage{mathptmx}

\usepackage{dcolumn}

\usepackage{tabularx}
\usepackage{array}

\usepackage{graphics}
\usepackage{graphicx}

\begin{document}

\section{Introduction}

When comparing experimental data with theoretical predictions on hadronic transitions, it is important to control the contribution of excited states. For example, light-cone sum rule determination for $g_{D^{*}D\pi}$ coupling failed to reproduce the experimental data unless one explicitly includes a negative contribution from the first radial excited state $D^{(*)'}$ state on the hadronic side of the sum rule \cite{BecirevicVP}.

The Generalized Eigenvalue Problem is a very efficient tool to deal with excited states on the lattice and can now be used with three-point correlation functions to extract matrix elements. We present a first estimate of the $g_{B^{*'}B\pi}$ coupling in the static limit of the Heavy Quark Effective Theory \cite{Eichten:1987xu}. Since the $B$ and $D$ mesons are degenerate in this limit, our result $g_{12} = -0.17(3)(2)_{\chi}$ is a first hint of the previous claim, the first error is statistical and the second originates from the chiral extrapolation. A more extensive discussion of the results will be found in the published paper \cite{Blossier:2013qma}.

\section{The $g_{B^{*'}B\pi}$ coupling}

The $g_{B^{*'}B\pi}$ coupling is defined by the following on-shell matrix element :
$$\big\langle B^{0}(p) \pi^{+}(q) | B^{*'+}(p',\epsilon^{(\lambda)}) \big\rangle = -  g_{B^{*'}B\pi}(q^2)  \times q_{\mu} \epsilon^{(\lambda) \mu}(p') \ .$$
\noindent Performing an LSZ reduction of the pion field and using PCAC relation, we are left with the following matrix element parametrized by three form factors :
\begin{align*}
\big\langle B^{*'+}(p',\epsilon^{(\lambda)}) | \mathcal{A}_{\mu} | B^{0}(p) \big\rangle &= 2m_{B^{*'}} A_0(q^2) \frac{\epsilon^{(\lambda)}\cdot q}{q^2}q^{\mu} + (m_B + m_{B^{*'}}) A_1(q^2) \left( \epsilon^{(\lambda) \mu} - \frac{\epsilon^{(\lambda)}\cdot q}{q^2} q^{\mu} \right) \\
& + A_2(q^2) \frac{\epsilon^{(\lambda)}\cdot q}{m_B+m_{B^{*'}}} \left[ (p_B+p_{B^{*'}})^{\mu} + \frac{m_B^2-m_{B^{*'}}^2}{q^2}q^{\mu} \right] \ .
\end{align*}
\noindent where $\mathcal{A}_{\mu}$ is the axial vector bilinear of light quarks and $B^{*'}$ is polarized in the $i$th direction. In the Heavy Meson Chiral Perturbation Theory (HM$\chi$PT) at leading order (static and chiral limit) and using the normalization of states $\langle B(\vec{p}) | B(\vec{p}) \rangle_{\mathrm{HQET}} = 1$, we just need to calculate $A_1(q^2_{\mathrm{max}})$ in the zero recoil kinematic configuration where $\vec{p}=\vec{p}'=\vec{0}$ and $q^2_{\mathrm{max}}=(m_{B^{*'}}-m_B)^2$. Choosing the quantization axis along the $z$ direction and the polarization vector $\epsilon^{(\lambda)}= (0,0,0,1)$ with the metric $(+,-,-,-)$, we define
$$g_{12} = \langle B^{*'}(\epsilon^{(\lambda)}) | \mathcal{A}_3 | B \rangle_{\mathrm{HQET}} \ \ \ , \ \ \ g_{12} = \frac{g_{B^{*'}B\pi}}{2\sqrt{m_B m_{B^{*'}}}} f_{\pi}$$

\section{Extracting the coupling from correlation functions}

We have to consider the following two-point correlation functions :
$$C^{(2)}_{\Pcal}(t) = \bigl\langle \ \sum_{\vec{y},\vec{x}} \Pcal(y) \Pcal^{\dag}(x) \ \big\rangle \big|_{y_0=x_0+t} \ \ \ ,\ \ \ C^{(2)}_{\Vcal}(t) =  \frac{1}{3} \sum_{i=1}^3 \bigl\langle \ \sum_{\vec{y},\vec{x}} \Vcal_i(y) \Vcal_i^{\dag}(x) \ \bigl\rangle \big|_{y_0=x_0+t} $$
\noindent where $\Pcal(x) = \sum_{y} \overline{h}(x) \gamma_5 \phi(x,y) \psi_l(y)$ and $\Vcal_i(x) = \sum_{y} \overline{h}(x) \gamma_i \phi(x,y) \psi_l(y)$ are respectively the heavy-light pseudoscalar and vector currents. But, due to the Heavy Quark Symmetry, they are equal and only one two-point correlation function has to be computed. We also need the following three-point correlation function~:
$$C_{ij}^{(3)}(t_z-t_x,t_y-t_x) = \big\langle \ \sum_{\vec{z},\vec{y},\vec{x}} \Vcal_3^{(i)}(z) \ \Acal_{3}(y) \ \Pcal^{(j) \dag}(x) \ \big\rangle \big|_{t_x<t_y<t_z} $$ 
\noindent where $\Acal_{\mu}= Z_{\Acal} \times \overline{\psi}_l(x) \gamma_{\mu}\gamma_5\psi_l(x)$ is the renormalized light-light axial current.\\

To deal with excited states, we have to solve generalized eigenvalue problems (GEVP) \cite{MichaelNE}-\cite{BlossierKD}. Since in the static limit of HQET pseudoscalar and vector meson are degenerate, we can actually solve just one GEVP~:
$$C^{(2)}(t) v_n(t,t_0) = \lambda_n(t,t_0) C^{(2)}(t_0) v_n(t,t_0)$$
\noindent where $C^{(2)}_{ij} = \langle \mathcal{O}_i(t) \mathcal{O}_j^{\dag}(0) \rangle$ is a $N\times N$ correlation matrix and $\mathcal{O}_i$ are interpolating fields with the correct quantum numbers. The sign of the eigenvectors is fixed by imposing the positivity of the decay constant $f_{B_n} = \langle B_n | \mathcal{O}_L | 0 \rangle$
where $\mathcal{O}_L$ refers to the local interpolating field. Then, we can construct ratios which tend toward the correct matrix element $g_{nm} = \langle B_n | A_3 | B^{*}_m \rangle$ at large time. We used two different methods, respectively called GEVP and sGEVP \cite{BulavaYZ} :
$$R^{\mathrm{GEVP}}_{mn}(t_2,t_1) = \frac{ \langle v_m(t_2,t_2-1) | C^{(3)}(t_1+t_2,t_1) | v_n(t_1,t_1-1) \rangle {\lambda_n(t_1+1,t_1)^{-t_1/2} {\lambda_m(t_2+1,t_2)^{-t_2/2}}  }}{Ê  \left(v_n(t_1,t_1-1),C^{(2)}(t_1)v_n(t_1,t_1-1)\right)^{1/2}     \left(v_m(t_2,t_2-1),C^{(2)}(t_2)v_m(t_2,t_2-1)\right)^{1/2}    }$$
$$R^{\mathrm{sGEVP}}_{mn}(t,t_0) = -\partial_t \left(  \frac{\left(v_m(t,t_0), \left[  K(t,t_0)/\lambda_n(t,t_0) - K(t_0,t_0) \right] v_n(t,t_0) \right)}{ \left(v_n(t,t_0),C(t_0)v_n(t,t_0)\right)^{1/2}   \left(v_m(t,t_0),C(t_0)v_m(t,t_0)\right)^{1/2} } e^{\Sigma(t_0,t_0) t_0/2} \right) $$
with
$$K_{ij}(t,t_0) = \sum_{t_1} e^{-(t-t_1)\Sigma(t,t_0)} C_{ij}^{(3)}(t,t_1) \ \ \ ,\ \ \  \Sigma(t,t_0) = E_n(t,t_0)-E_m(t,t_0)$$
\noindent where $(a,b)=\sum_i a_i b_i$. These ratios converge quickly to the desired coupling constant as the contribution of higher excited states are strongly suppressed \cite{BulavaYZ} \cite{Blossier:2013qma}:
\begin{align*}
R_{mn}^{\mathrm{GEVP}}   &\xrightarrow{t_1\gg 1, t_2\gg 1}  g_{nm} + \mathcal{O}\left( e^{-\Delta_{N+1,m}t_1} , e^{-\Delta_{N+1,n}t_2}   \right) \\[2mm]
R_{mn}^{\mathrm{sGEVP}} &\xrightarrow[\ \ t_0=t-1 \ \ ]{t\gg 1}  g_{nm} + \mathcal{O}\left( t e^{-\Delta_{N+1,n}t} \right) \ \ \ n<m\\
					  &\xrightarrow[\ \ t_0=t-1 \ \ ]{t\gg 1}  g_{nm} + \mathcal{O}\left( e^{-\Delta_{N+1,m}t}  \right) \ \ \ n>m
\end{align*} 
where $\Delta_{N+1,m} = E_{N+1} - E_m$ and $N$ is the size of the GEVP. In the following, we choose $t_1=t_2$. Since $t=t_1+t_2$, we expect a faster suppression of higher excited states in the case of the sGEVP.

\section{Lattice setup}

To perform our lattice computation, we used $N_f=2$ gauge configurations from CLS ensembles with different pion masses ($310~\mev \leq m_{\pi} \leq 440~\mev$) and three lattice spacings ($0.05~\fm \lesssim a \lesssim 0.08~\fm$). The details of the configurations analyzed in this work are listed in table \ref{lattice_setup}. These simulations use non-perturbatively $\mathcal{O}(a)$-improved Wilson quarks and the HYP2 discretization for the static quark action \cite{HasenfratzHP} \cite{DellaMorteYC}. Correlation functions are estimated using all-to-all light quark propagators with full time dilution \cite{FoleyAC}.
\begin{table}[h!]
\begin{center}
\begin{tabular}{|c|c|c|c|c|c|c|}
\hline
CLS label	&	$\beta$	&	$L^3\times T$		&	$\kappa$	&	a [fm]	&	$m_\pi$ [MeV]	&	\#\\
\hline
A5		&	5.2		&	$32^3\times 64$	&	0.13594	&	0.075	&	330			&	500\\
\hline
E5		&	5.3		&	$32^3\times 64$	&	0.13625	&	0.065	&	440			&	500\\
F6		&			&	$48^3 \times 96$	&	0.13635	&			&	310			&	600\\
\hline
N6		&	5.5		&	$48^3 \times 96$	&	0.13667	&	0.048	&	340			&	400\\
\hline
\end{tabular}
\end{center}
\caption{Parameters of the simulations.}
\label{lattice_setup} 
\end{table} 
\noindent We used $N=4$ interpolating fields of the Gaussian smeared-form $\mathcal{O}^{(i)} = \overline{h} \gamma_5 (1+\kappa_G a^2 \Delta )^{R_i} \psi_l$  \cite{GuskenAD} where $\kappa_G=0.1$, $r_i=2a\sqrt{\kappa_G R_i} \leq 0.6~\fm$ and $\Delta$ is a gauge covariant Laplacian made of three times APE-blocked links \cite{AlbaneseDS}. The axial current renormalisation constant $Z_{\Acal}$ was determined non perturbatively by the ALPHA collaboration in \cite{DellaMorteXB}, \cite{FritzschWQ} and the scale was set through the kaon decay constant \cite{Fritzsch:2012wq}. Statistical errors are estimated from a jackknife procedure.

\section{Results}

\begin{figure}[h!]
\centering 
\includegraphics*[width=7cm, height=5cm]{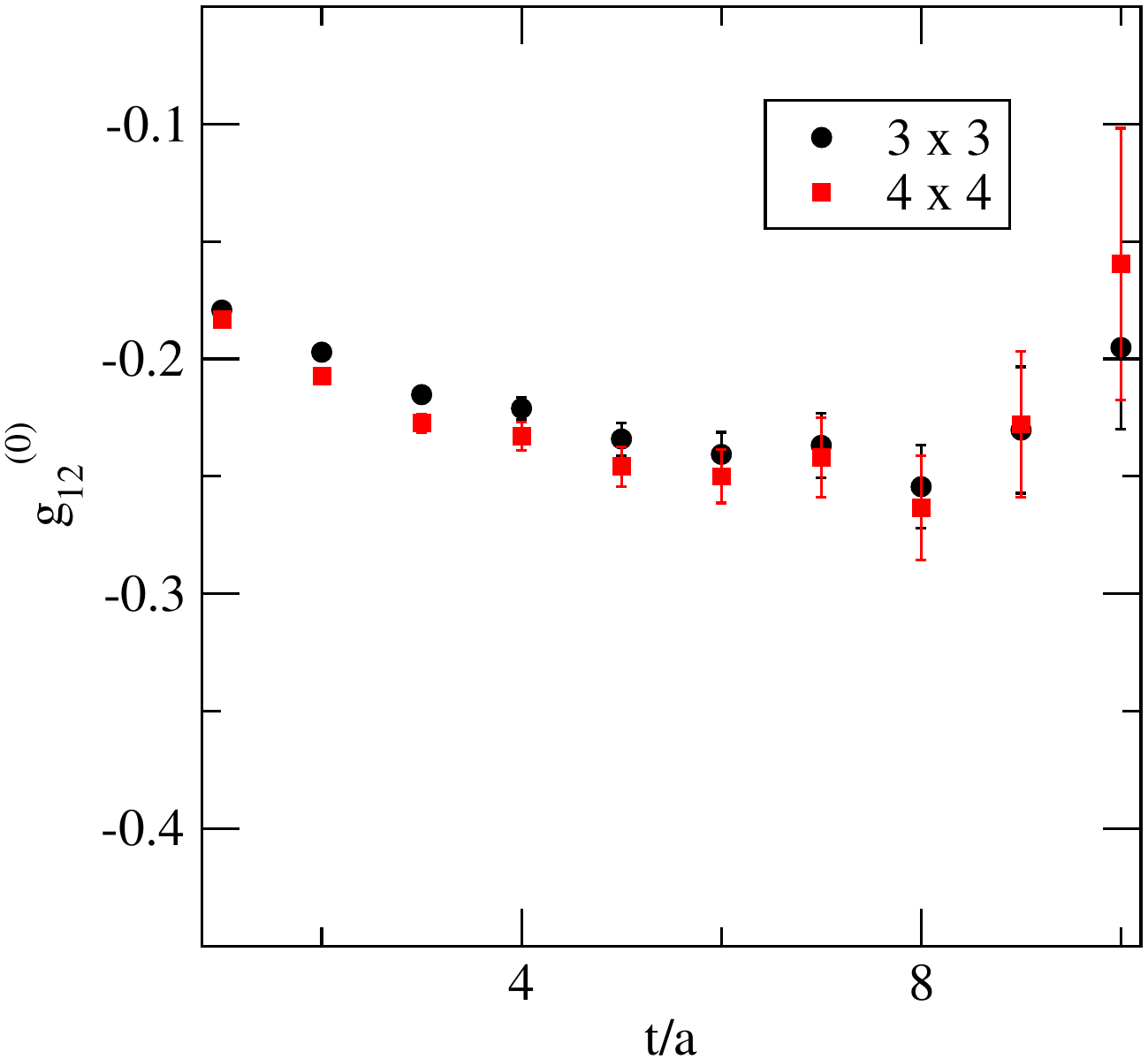} \quad
\includegraphics*[width=7cm, height=5cm]{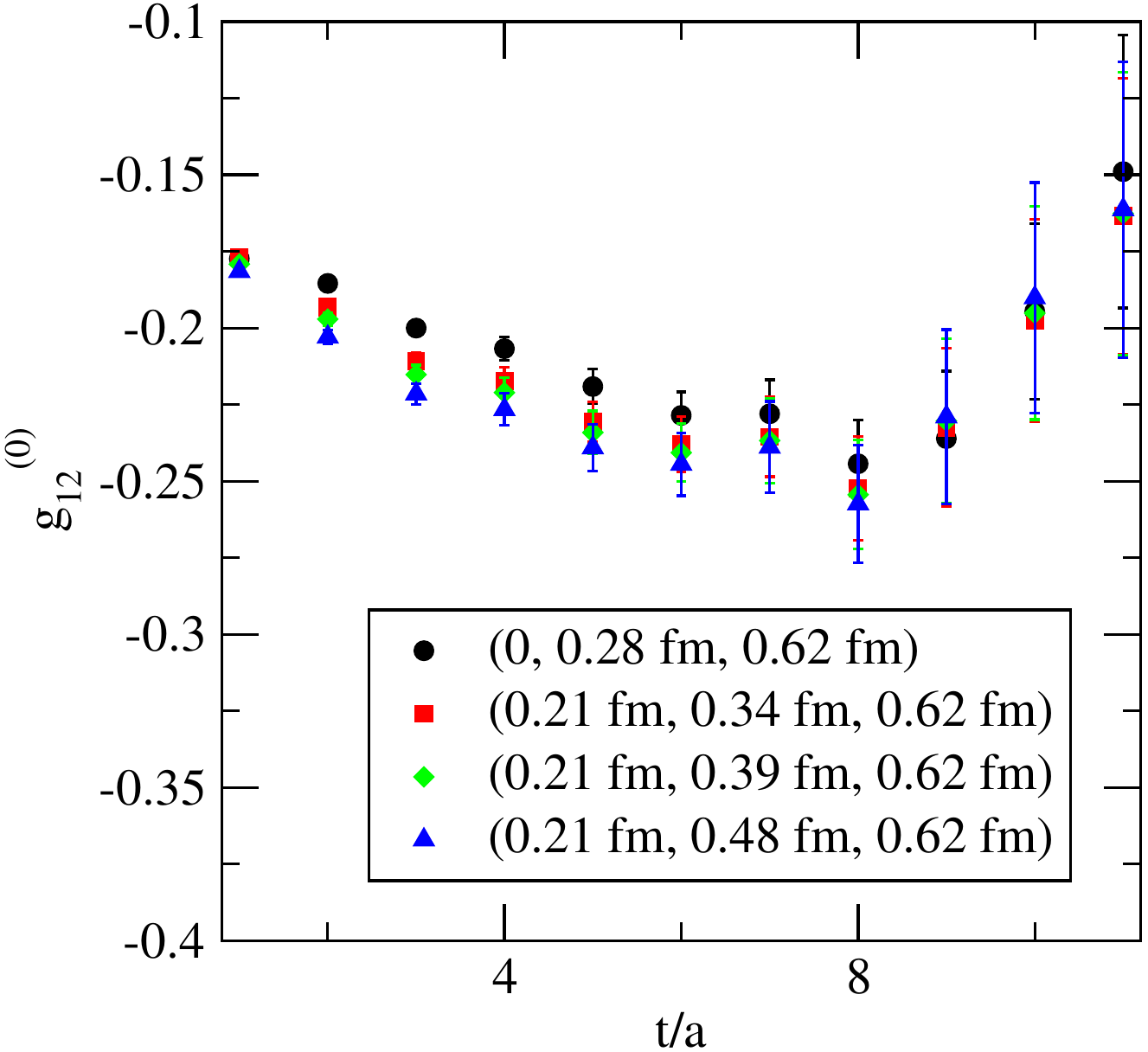}
\caption{Dependence of bare $g_{12}$ on the size of the GEVP (left) and on the radius of wave functions (right) for the CLS ensemble E5.}
\label{GEVP_stability}
\end{figure}

To check the stability of our results, we have solved both $3\times3$ and $4\times4$ GEVP and tested different combinations of interpolating fields, results are shown in Figure \ref{GEVP_stability}.

\begin{figure}[t]
\includegraphics*[width=7cm, height=5cm]{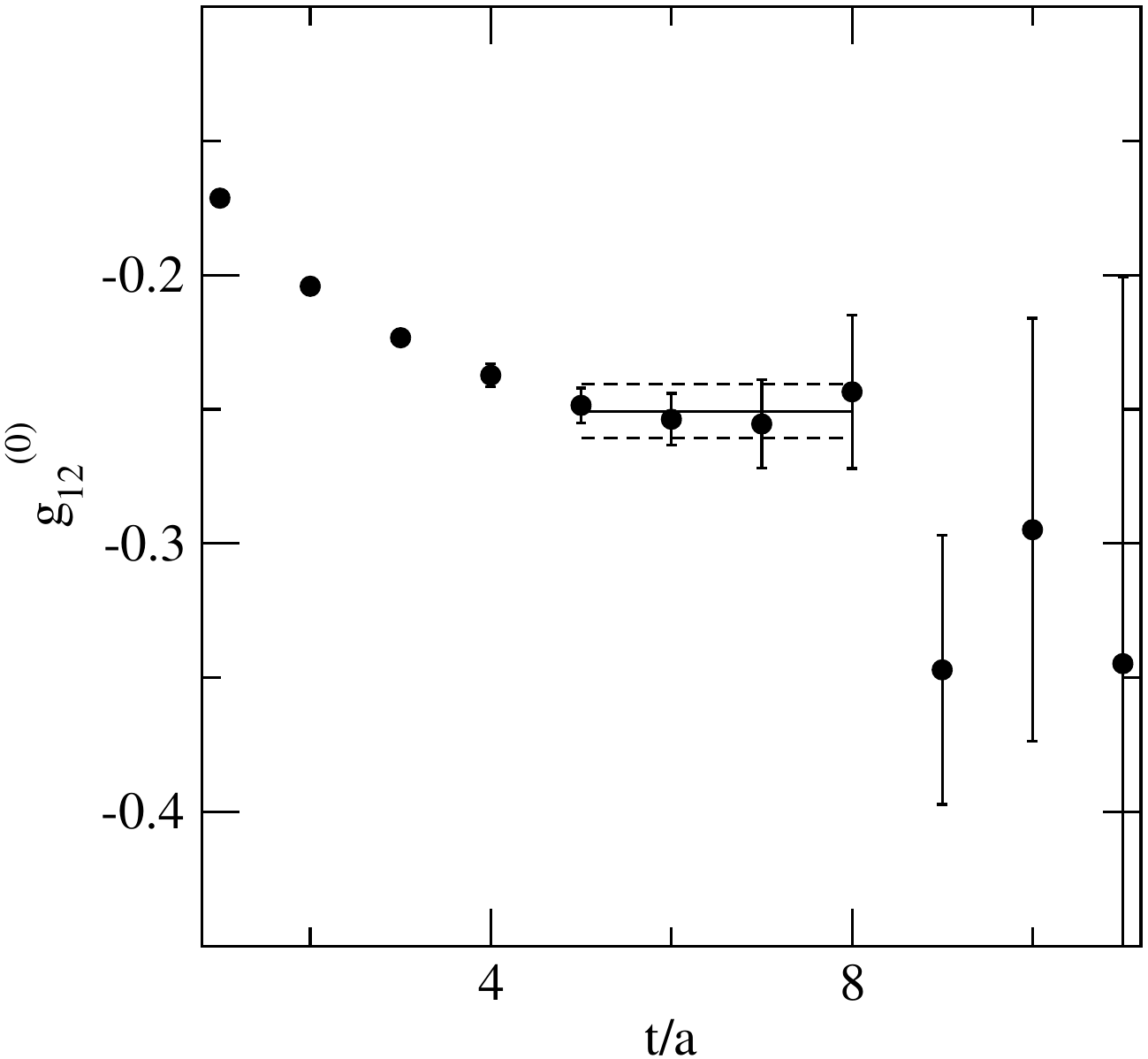} \quad
\includegraphics*[width=7cm, height=5cm]{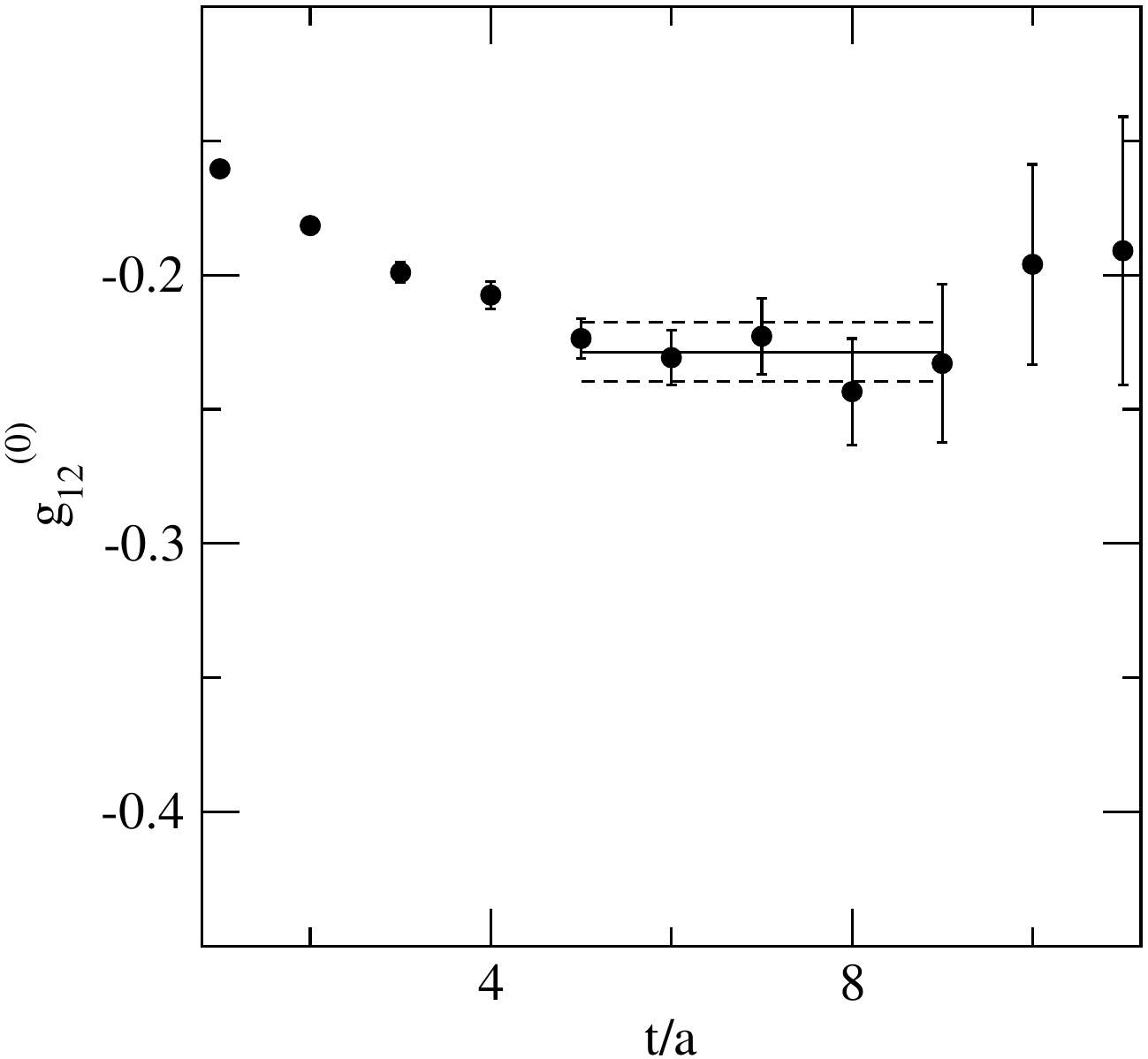}
\caption{Plateaus of bare $g_{12}$ extracted by GEVP (left) and sGEVP (right) for the CLS ensemble E5.}
\label{gevp_vs_sgevp} 
\end{figure}

\begin{figure}[h!]
\centering
\includegraphics*[width=7cm, height=6cm]{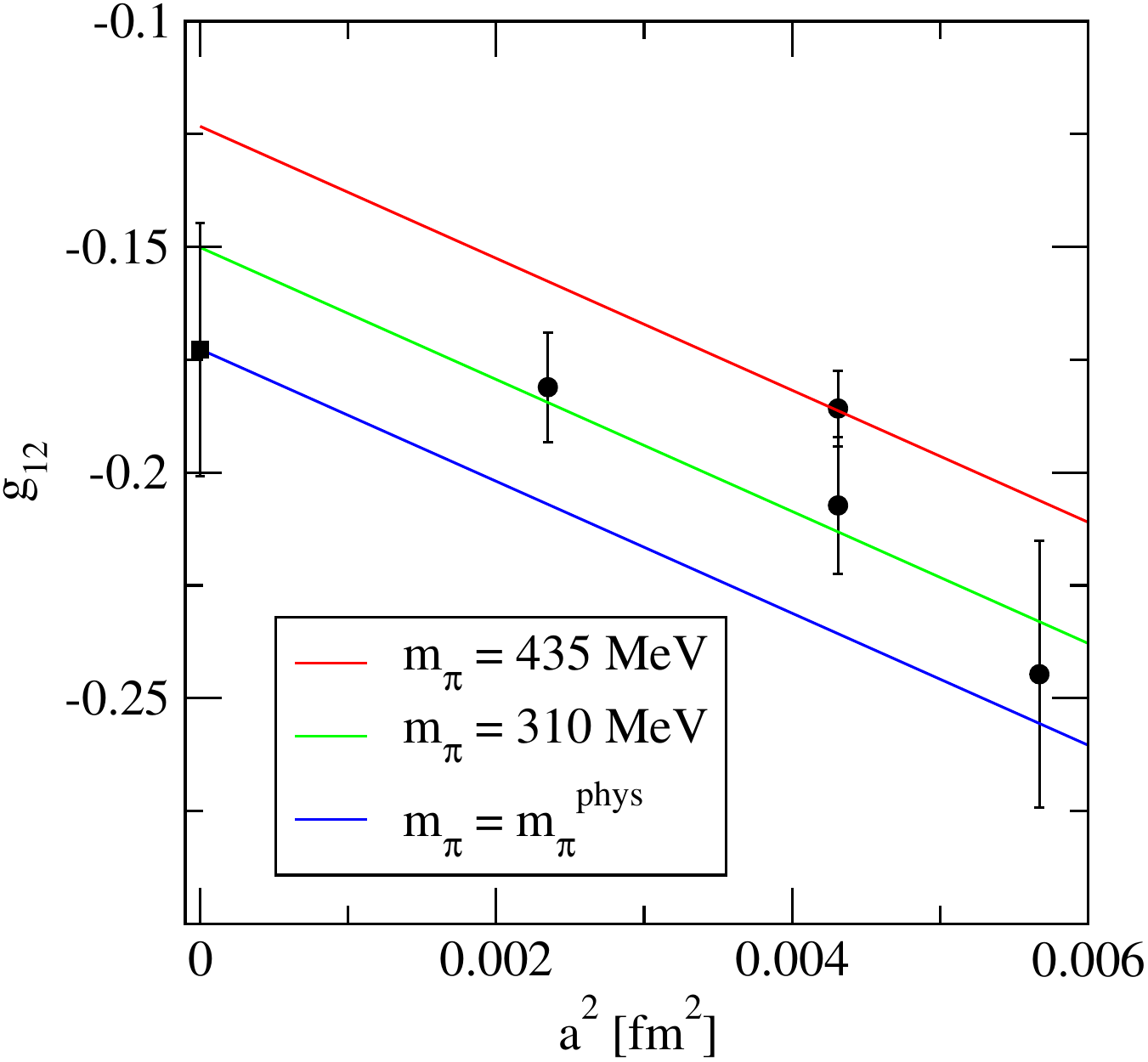}
\caption{Continuum and chiral extrapolation of $g_{12}$.}
\label{g12_extrapolaltion}
\end{figure}

\noindent Moreover, as shown in figure \ref{gevp_vs_sgevp} both GEVP and sGEVP results are consistent, but with a better behavior at large time in the case of the sGEVP. Therefore, the value of the coupling for each ensemble in Table \ref{res_g} and in the following, corresponds to the sGEVP only. Inspired by Heavy Meson Chiral Perturbation Theory \cite{Casalbuoni:1996pg} \cite{Burdman:1992gh} and due to the fact that our action and correlations functions are $\mathcal{O}(a)$ improved, we tried two fit formulae for the extrapolation to the physical point :
\begin{align}
g_{12}&=C_0 + a^2 C_1\,,  \label{fit1} \\
g_{12}&=C'_0 + a^2 C'_1 + m^2_\pi C'_2\,.  \label{fit2}
\end{align}
\noindent We show in Figure \ref{g12_extrapolaltion} the continuum and chiral extrapolations. Since the two fits are consistent, we used the result (\ref{fit2}) as central value and obtain :
$$g_{12} = -0.17(3)(2)_{\chi}$$
where the first error is statistical and the second originates from the chiral extrapolation and is estimated as the discrepancy between (\ref{fit1}) and (\ref{fit2}). Fit parameters are collected in Table \ref{res_g}.

\begin{figure}[h!]
\centering 
\includegraphics*[width=7cm, height=5cm]{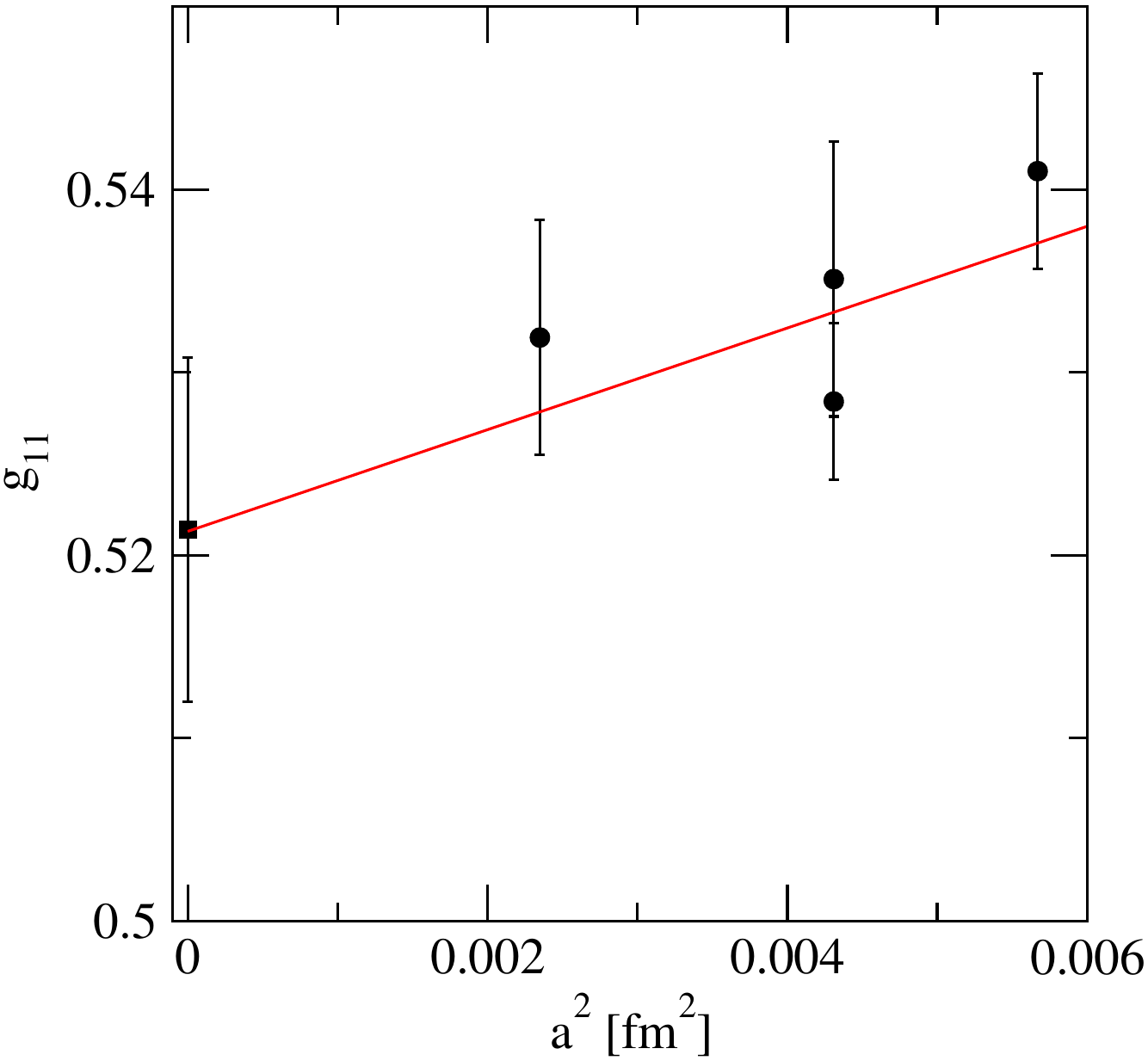} \quad
\includegraphics*[width=7cm, height=5cm]{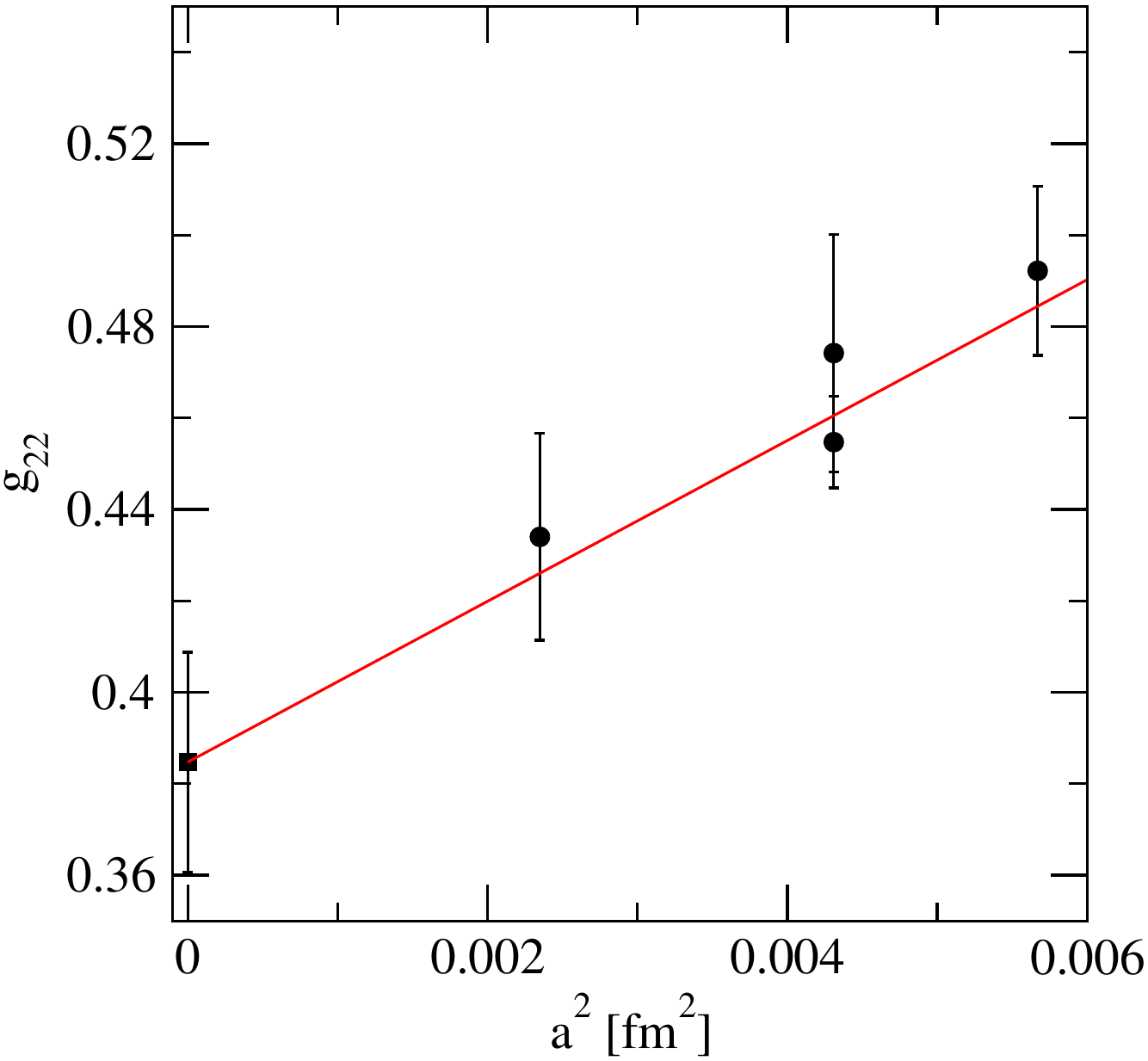}
\caption{Extrapolation to the continuum and chiral limit of $g_{11}$ and $g_{22}$}
\label{g11_g22}
\end{figure}

\noindent Finally, we have to check that we are safe from multi-hadron thresholds due to the emission of pions. The P-wave decay $B^{*'} \rightarrow B^{*}(\vec{p}) \pi(-\vec{p})$ is kinematically forbidden since $L < 3~\fm$. The second, potentially dangerous, decay is the S-wave decay $B^{*'} \rightarrow B_1^{*} \pi$. But, examining our lattice results for $\Sigma_{12}$, listed in Table \ref{res_g}, we have $230~\mathrm{MeV} \leq m_{B^{*'}} - m_B - m_{\pi} \leq 360~\mathrm{MeV}$. Then using recent lattice results \cite{Michael:2010aa} with similar lattice spacings : $400~\mathrm{MeV} \leq m_{B_1^{*}} - m_B \leq 500~\mathrm{MeV}$, we can conclude that this decay is also forbidden. Finally, as a byproduct of our calculation, we also obtain $g_{11}=0.52(2)$, in excellent agreement with a computation by the ALPHA collaboration focused on that quantity \cite{BulavaEJ}, and $g_{22}=0.38(4)$. The continuum and chiral extrapolations for these quantities are shown in Figure \ref{g11_g22} while the value of these coupling for each ensemble are listed in Table \ref{res_g}.

\begin{table}
\begin{center}
\begin{tabular}{|c|c|c|c|c|}
\cline{2-5}
\multicolumn{1}{l|}{}&$a\Sigma_{12}$&$g_{12}$&$g_{11}$&$g_{12}$\\
\hline
A5	&	0.255(8)	&	-0.245(29) 	&	0.541(5)	&	0.492(19)\\
E5	&	0.222(8)	&	-0.186(8)		&	0.535(8)	&	0.455(10)\\
F6	&	0.216(12)	&	-0.207(15)		&	0.528(4)	&	0.474(26)\\
N6	&	0.173(7)	&	-0.181(12)		&	0.532(6)	&	0.434(23)\\
\hline
\end{tabular}
\quad
\begin{tabular}{|c|c|c|}
\cline{2-3}
\multicolumn{1}{l|}{}&fit (\ref{fit1})&fit (\ref{fit2})\\
\hline
$C_{0}$&-0.178(29)&-0.155(26)\\
$C_{1}$&-14.6(7.3)&-9.2(6.6)\\
$C_{2}$&0.29(16)&-\\
\hline
\end{tabular}
\end{center}
\caption{Value of the mass splitting $s\Sigma_{12}$ in lattice units and $g_{12}$ for the different ensembles (left) and fit parameters of eq. (\protect\ref{fit1}) and (\protect\ref{fit2}) (right).}
\label{res_g}
\end{table}

\section{Conclusion}

We have performed a first estimate of the axial form factor $A_1(q^2_{\mathrm{max}})=g_{12}$ parametrizing the decay $B^{*'}\rightarrow B$ at zero recoil and in the static limit of HQET from $N_f = 2$ lattice simulations. We have obtained a negative value for this form factor. It is almost three times smaller than the $g_{11}$ coupling, but not compatible with zero : $g_{12}=-0.17(4)$ while $g_{11}=0.52(2)$. Moreover we find $g_{22} = 0.38(4)$, which is not strongly suppressed with respect to $g_{11}$. Our work is a first hint of confirmation of the statement made in \cite{BecirevicVP} to explain the small value of $g_{D^{*}D\pi}$ computed analytically when compared to experiment. 

\section*{Acknowledgements}

This work was granted access to the HPC resources of CINES under the allocations 2012-056808 and 2013-056808 made by GENCI



\begin{thebibliography}{99}

\bibitem{BecirevicVP}
  D.~Becirevic, J.~Charles, A.~LeYaouanc, L.~Oliver, O.~Pene and J.~C.~Raynal,
JHEP {\bf 0301}, 009 (2003).
[hep-ph/0212177].

\bibitem{BulavaEJ}
  J.~Bulava {\it et al.}  [ALPHA Collaboration],
PoS LATTICE {\bf 2010}, 303 (2010),
[arXiv:1011.4393 [hep-lat]]; 
J. Bulava {\it et al.}  [ALPHA Collaboration], in preparation.

\bibitem{BecirevicPF}
  D.~Becirevic and F.~Sanfilippo,
[arXiv:1210.5410 [hep-lat]].

\bibitem{Eichten:1987xu}
  E.~Eichten,
  Nucl.\ Phys.\ Proc.\ Suppl.\  {\bf 4} (1988) 170.
  
\bibitem{KhodjamirianHB}
  A.~Khodjamirian, R.~Ruckl, S.~Weinzierl and O.~I.~Yakovlev,
Phys.\ Lett.\ B {\bf 457}, 245 (1999).
[hep-ph/9903421].

\bibitem{BulavaNP}
  J.~Bulava,
PoS LATTICE {\bf 2011}, 021 (2011).
[arXiv:1112.0212 [hep-lat]].

\bibitem{BurchCC}
  T.~Burch, C.~Gattringer, L.~Y.~Glozman, C.~Hagen, D.~Hierl, C.~B.~Lang and A.~Schafer,
Phys.\ Rev.\ D {\bf 74}, 014504 (2006).
[hep-lat/0604019].


\bibitem{BlossierVZ}
  B.~Blossier {\it et al.}  [ALPHA Collaboration],
JHEP {\bf 1005}, 074 (2010).
[arXiv:1004.2661 [hep-lat]].


\bibitem{MohlerKE}
  D.~Mohler and R.~M.~Woloshyn,
Phys.\ Rev.\ D {\bf 84}, 054505 (2011).
[arXiv:1103.5506 [hep-lat]].

\bibitem{MahbubRM}
  M.~S.~Mahbub {\it et al.}  [CSSM Lattice Collaboration],
Phys.\ Lett.\ B {\bf 707}, 389 (2012).
[arXiv:1011.5724 [hep-lat]].

\bibitem{MichaelNE}
  C.~Michael,
Nucl.\ Phys.\ B {\bf 259}, 58 (1985)..

\bibitem{LuscherCK}
  M.~Luscher and U.~Wolff,
Nucl.\ Phys.\ B {\bf 339}, 222 (1990)..

\bibitem{BlossierKD}
  B.~Blossier, M.~Della Morte, G.~von Hippel, T.~Mendes and R.~Sommer,
JHEP {\bf 0904}, 094 (2009).
[arXiv:0902.1265 [hep-lat]].


\bibitem{BulavaYZ}
  J.~Bulava, M.~Donnellan and R.~Sommer,
JHEP {\bf 1201}, 140 (2012).
[arXiv:1108.3774 [hep-lat]].

\bibitem{HasenfratzHP}
  A.~Hasenfratz and F.~Knechtli,
Phys.\ Rev.\ D {\bf 64}, 034504 (2001).
[hep-lat/0103029];

\bibitem{DellaMorteYC}
 M.~Della Morte, A.~Shindler and R.~Sommer,
JHEP {\bf 0508}, 051 (2005).
[hep-lat/0506008].


\bibitem{FoleyAC}
  J.~Foley, K.~Jimmy Juge, A.~O'Cais, M.~Peardon, S.~M.~Ryan and J.~-I.~Skullerud,
Comput.\ Phys.\ Commun.\  {\bf 172}, 145 (2005).
[hep-lat/0505023].

\bibitem{GuskenAD}
  S.~Gusken, U.~Low, K.~H.~Mutter, R.~Sommer, A.~Patel and K.~Schilling,
Phys.\ Lett.\ B {\bf 227}, 266 (1989)..

\bibitem{AlbaneseDS}
  M.~Albanese {\it et al.}  [APE Collaboration],
Phys.\ Lett.\ B {\bf 192}, 163 (1987)..

  \bibitem{Blossier:2013qma}
  B.~Blossier, J.~Bulava, M.~Donnellan and A.~G\'erardin,
  Phys.\ Rev.\ D {\bf 87} (2013) 094518
  [arXiv:1304.3363 [hep-lat]].
  
  
 \bibitem{Fritzsch:2012wq}
  P.~Fritzsch, F.~Knechtli, B.~Leder, M.~Marinkovic, S.~Schaefer, R.~Sommer, F.~Virotta and R.~Sommer {\it et al.},
  Nucl.\ Phys.\ B {\bf 865} (2012) 397
  [arXiv:1205.5380 [hep-lat]].
  
\bibitem{DellaMorteXB}
  M.~Della Morte, R.~Sommer and S.~Takeda,
Phys.\ Lett.\ B {\bf 672}, 407 (2009).
[arXiv:0807.1120 [hep-lat]].

\bibitem{FritzschWQ}
  P.~Fritzsch, F.~Knechtli, B.~Leder, M.~Marinkovic, S.~Schaefer, R.~Sommer and F.~Virotta,
Nucl.\ Phys.\ B {\bf 865}, 397 (2012).
[arXiv:1205.5380 [hep-lat]].

\bibitem{Michael:2010aa}
  C.~Michael {\it et al.}  [ETM Collaboration],
  JHEP {\bf 1008} (2010) 009
  [arXiv:1004.4235 [hep-lat]].

\bibitem{Casalbuoni:1996pg}
  R.~Casalbuoni, A.~Deandrea, N.~Di Bartolomeo, R.~Gatto, F.~Feruglio and G.~Nardulli,
  Phys.\ Rept.\  {\bf 281} (1997) 145
  [hep-ph/9605342].
  
  \bibitem{Burdman:1992gh}
  G.~Burdman and J.~F.~Donoghue,
  Phys.\ Lett.\ B {\bf 280} (1992) 287.
  
  
\end{thebibliography}
\end{document}